\documentclass[10pt,pra,aps,twocolumn,superscriptaddress,floatfix,notitlepage,nofootinbib,reprint]{revtex4-2}
\usepackage{amsmath}
\usepackage{amsfonts}
\usepackage{amssymb}
\usepackage[pdftex]{graphicx}
\usepackage{hyperref}
\usepackage{xcolor}
\usepackage{epsdice}
\usepackage{natbib}
\usepackage{physics}
\usepackage[inkscapelatex=false]{svg}

\begin{document}
\title{Quantifying nonclassicality of mixed Fock states}
\author{Spencer Rogers}
\email{spencer.rogers@uri.edu}
\affiliation{Department of Physics, University of Rhode Island, Kingston, RI 02881, USA}

\author{Tommy Muth}
\affiliation{Department of Physics, University of Rhode Island, Kingston, RI 02881, USA}

\author{Wenchao Ge}
\email{wenchao.ge@uri.edu}
\affiliation{Department of Physics, University of Rhode Island, Kingston, RI 02881, USA}

\date{\today}
\begin{abstract}
Nonclassical states of bosonic modes are important resources for quantum-enhanced technologies. Yet, quantifying nonclassicality of these states, in particular mixed states, can be a challenge. Here we present results of quantifying the nonclassicality of a bosonic mode in a mixed Fock state via the operational resource theory (ORT) measure [W. Ge, K. Jacobs, S. Asiri, M. Foss-Feig, and M. S. Zubairy, Phys. Rev. Res. 2, 023400 (2020)], which relates nonclassicality to metrological advantage. Generally speaking, evaluating a resource-theoretic measure for mixed states is challenging, since it involves finding a convex roof. However, we show that our problem can be reduced to a linear programming problem. By analyzing the results of numerical optimization, we are able to extract analytical results for the case where three or four neighboring Fock states have nonzero population. Interestingly, we find that such a mode can be in distinct phases, depending on the populations. Lastly, we demonstrate how our method is generalizable to density matrices of higher ranks. Our findings suggest a viable method for evaluating nonclassicality of arbitrary mixed bosonic states and potentially for solving other convex roof optimization problems.
\end{abstract}
\maketitle

\section{Introduction}
For bosonic modes in quantum optics and other analogous fields, the coherent states, defined as eigenstates of the annihilation operator, are regarded as the most classical~\cite{Hillery:1985aa}. Not only do they have minimal uncertainty and maintain their shape under the harmonic oscillator Hamiltonian, but also they remain separable from each other when subject to linear interactions (as occur in beam-splitters). By contrast, superpositions of coherent states (e.g., cat and squeezed states) may have highly nonclassical features, and with these the potential to revolutionize such technologies as communication \cite{braunstein1998teleportation}, computation \cite{braunstein2005quantum,mirrahimi2014dynamically}, and metrology \cite{giovannetti2006quantum}. Indeed, on the metrological side, squeezed states have already become an important tool in advanced gravitational wave interferometry \cite{aasi2013enhanced,tse2019quantum}.

Given the applications of nonclassicality, quantifying it, i.e., evaluating the nonclassicality of different bosonic states, is an important problem. Various measures of nonclassicality have been proposed, including the nonclassicality depth \cite{lee1991measure}, nonclassical distance \cite{marian2002quantifying, hillery1987nonclassical}, quantifications in terms of the Schmidt rank \cite{gehrke2012quantification,vogel2014unified}, and resource-theoretic measures \cite{tan2017quantifying, YadinPRX18, Kwon19}. These measures satisfy the basic criteria that they are non-negative, and zero only for classical states (i.e., coherent states or, if mixed, classical mixtures of coherent states). They also characterize certain aspects of coherent superposition, such as the minimum number of coherent superpositions in a quantum state (in the case of the Schmidt rank) \cite{gehrke2012quantification}.

Here, we focus on the operational resource theory (ORT) measure of nonclassicality \cite{ge2020operational, ge2020evaluating}, which has two defining features. First, it is \textit{operational}, since it is related to the usefulness of a state for certain tasks. For pure states, the ORT measure directly quantifies the metrological power of a state in terms of the quantum Fisher information of quadrature sensing~\cite{Ge2023njp}, while for mixed states, the ORT measure is a tight upper bound on this power; this is the closest possible relationship because some nonclassical mixed states have zero metrological advantage over coherent states. This feature endows the ORT measure an operational meaning to evaluate different nonclassical states on the same footing. Second, the ORT measure is \textit{resource-theoretic}. A quantum resource theory~\cite{RevModPhys.91.025001} defines free states and resource states via the operations, called \textit{free operations}, under which it is impossible to increase that resource. In the ORT measure of nonclassicality, coherent states, and classical mixtures of them, are considered the free states, while nonclassical states are considered resource states. The free operations are classical operations (meaning operations that cannot create superpositions of coherent states from mixtures of coherent states) --- Ref. \cite{ge2020operational} rigorously showed that the ORT measure does not increase under such operations. In addition, the ORT measure has an interesting feature of quantifying the size, or macroscopicity~\cite{Frowis2018}, of a coherent superposition state~\cite{ge2020operational}

Previous works computed the ORT measure, from here on also referred to as the \textit{nonclassicality} $\mathcal{N}$, for various pure bosonic states of interest \cite{ge2020operational,ge2020evaluating,merlin2022operational}. It has been shown, for example, that squeezed vacuum has the highest nonclassicality per unit average energy (which speaks to the metrological usefulness of squeezed vacuum) \cite{ge2020operational}. Cat states whose nonclassicality per unit energy approach that of squeezed vacuum in the asymptotic limit of large energy have also been discovered \cite{ge2020evaluating}, thus extending our knowledge of optimal sensing with Mach-Zehnder interferometry~\cite{Lang13,ge2020operational}.

While pure states are ideal for applications, \textit{mixed} states are inevitable in practice, due to coupling to the environment. For example, mixtures of Fock states can be generated under a dephasing channel in various bosonic systems~\cite{Ferrini10,Arqand2020,Strzalka2024}. Computing a resource-theoretic measure for a mixed state $\hat{\rho}$, however, poses a significant challenge. This is because the ORT measure for mixed states is a convex roof construction, similar to entanglement measures for mixed states \cite{toth2015evaluating, chen2005entanglement} --- computing it involves extremizing a function over all the decompositions of $\hat{\rho}$. Previous calculations only address very simple classes of states \cite{ge2020evaluating}. For more general states, only a lower bound has been given: $\mathcal{N}(\hat{\rho})\geq\mathcal{W}(\hat{\rho})$, where $\mathcal{W}(\hat{\rho})$ is the metrological power (to be defined later) \cite{ge2020evaluating}. In this work, we present a numerical method of linear programming for calculating the ORT measure for mixed Fock states. Based on the numerical results, we extract analytical ansatzes for the ORT measure of rank-3 and rank-4 mixed states and identify distinct phases depending on the populations. Our method provides a viable way to evaluate arbitrary mixed Fock states of bosonic systems.

Our paper is structured as follows: in section II, we review the ORT measure of nonclassicality for pure and mixed states. In section III, we explain our method, based on linear programming, for calculating the ORT measure for mixed states. We focus on one class of mixed states in particular: those that are diagonal in the Fock basis. In section IV, we apply our method to rank-3 and rank-4 states, and show that such states can be in various phases, depending on their populations. The results here are largely analytical, as were able to form analytical ansatzes based on the numerical results. For higher-rank states, numerical results are possible, as we demonstrate on truncated thermal states. Section V contains our concluding remarks.

\section{The ORT Measure}
\subsection{Pure states}
A pure bosonic state $\ket{\psi}$ is defined as \textit{nonclassical} if and only if it is not a coherent state, i.e., $\hat{a}\ket{\psi}\neq\alpha\ket{\psi}$ for all complex numbers $\alpha$~\cite{Hillery:1985aa}. The ORT measure $\mathcal{N}$ assigns a nonclassicality value to each pure state as~\cite{ge2020operational,ge2020evaluating}:
\begin{equation}\label{eq:ORTMeasurePure}
\begin{split}
\mathcal{N}(\ket{\psi})&=\max_\mu\bra{\psi}(\Delta \hat{X}_\mu)^2\ket{\psi}-\frac{1}{2}\\
&=\langle\hat{a}^\dagger\hat{a}\rangle-\left|\langle\hat{a}\rangle\right|^2+\left|\langle\hat{a}^2\rangle-\langle\hat{a}\rangle^2\right|.
\end{split}
\end{equation}
Here, $(\Delta \hat{X}_\mu)^2$ is the variance of a quadrature $\hat{X}_\mu=i(e^{-i\mu}\hat{a}^\dagger-e^{i\mu}\hat{a})/\sqrt{2}$, $\hat{a}$ is the annihilation operator for the mode and $\mu\in[0,2\pi]$.

Whereas the second line in Eq. ~\eqref{eq:ORTMeasurePure} is generally more convenient for calculations, the first line makes the connection to metrological power more apparent. In quantum metrology with unitary encoding, one measures a classical parameter $\theta$ by applying a transformation $\hat{U}(\theta)=e^{-i\theta\hat{G}}$ to a quantum sensor, where the generator $\hat{G}$ is some Hermitian observable of the quantum sensor. $\hat{U}(\theta)$ may represent the effect of coupling the quantum sensor to a classical system, and may or may not describe some interaction picture. Regardless, the effective Hamiltonian for the sensor is $\hat{G}$, while $\theta$ acts analogously to time. Assuming the sensor is prepared in a pure state $\ket{\psi}$, the energy-time uncertainty principle tells us that, the larger the variance $\bra{\psi}(\Delta\hat{G})^2\ket{\psi}$, the faster the sensor will evolve with respect to $\theta$. So, the larger the variance $\bra{\psi}(\Delta\hat{G})^2\ket{\psi}$, the more responsive the sensor will be to small variations in $\theta$. Indeed, the quantum Fisher information (QFI)~\cite{Braunstein94} for this protocol, which quantifies the precision with which $\theta$ can be measured, is simply this variance $F_{\hat{G}}(\ket{\psi})=\bra{\psi}(\Delta\hat{G})^2\ket{\psi}$.\footnote{In this paper, we drop a factor of $4$ from the standard definition of the QFI.} From the first line of Eq. ~\eqref{eq:ORTMeasurePure}, we see that the ORT measure for pure states is based on the maximum quadrature variance --- in other words, the Fisher information when using the sensor's optimal quadrature as the generator. Since the maximum quadrature variance of a coherent state is always $\frac{1}{2}$, $\mathcal{N}(\ket{\alpha})=0$ for coherent states $\ket{\alpha}$. For nonclassical states, the maximum quadrature variance is always greater than $\frac{1}{2}$, and $\mathcal{N}>0$ describes the sensing enhancement over coherent states. We refer to this sensing enhancement as the \textit{metrological power} $\mathcal{W}$~\cite{ge2020operational, Ge2023njp}. For pure states, $\mathcal{N}(\ket{\psi})=\mathcal{W}(\ket{\psi})$.

One may wonder why the ORT measure for pure states is based only on the maximum variance \textit{quadrature}, as opposed to the maximum variance Hermitian generator $\hat{G}$ in general (quadrature or not). This has to do with the resource-theoretic nature of the ORT measure. A unitary $e^{-i\theta\hat{X}_\mu}$ is a free operation for classical states~\cite{tan2017quantifying,ge2020operational}--- when it acts on a coherent state, it displaces it in phase space, yielding a new coherent state. Another free operator for evaluating nonclassicality is to use a phase shifter, i.e., $\hat{G}=\hat{a}^{\dagger}\hat{a}$. It is shown that the generator turns into an effective quadrature operator $\hat{X}_\mu$ in a multi-port phase sensing scheme with a single-mode nonclassical input in combination with classical light~\cite{Ge2023njp}. 

\subsection{Mixed states}

A mixed bosonic state is defined as classical if and only if its density matrix $\hat{\rho}$ can be regarded as a classical mixture of coherent states. Otherwise, the state is considered nonclassical. Equivalently, one may determine if a state is classical by examining its Glauber-Sudarshan $P$-function. Any bosonic state $\hat{\rho}$ can be represented using its Glauber-Sudarshan $P$-function \cite{Glauber63,Sudarshan63} as
\begin{equation}\label{eq:GlauberSudarshanPfunction}
\hat{\rho}=\int P(\alpha,\alpha^*)\ket{\alpha}\bra{\alpha}d^2\alpha.
\end{equation}
The state $\hat{\rho}$ is defined as classical if $P(\alpha,\alpha^*)$ is positive semidefinite, in which case $P(\alpha,\alpha^*)$ acts as a classical probability density over the coherent states $\ket{\alpha}$. Otherwise, the state is nonclassical \cite{SZ,lee1991measure}.

The ORT measure of nonclassicality for mixed states is~\cite{ge2020operational}:
\begin{equation}\label{eq:ORTMeasureMixed}
\begin{split}
\mathcal{N}(\hat{\rho})&=\min_{\{q_j,\ket{\phi_j}\}}\left[\max_\mu\sum_jq_j\bra{\phi_j}(\Delta \hat{X}_\mu)^2\ket{\phi_j}\right]-\frac{1}{2}\\
&=\min_{\{q_j,\ket{\phi_j}\}}\left[\sum_jq_j\left(\bar{n}_j-|\Bar{\alpha}_j|^2\right)+\left|\sum_jq_j\left(\Bar{\xi}_j-\Bar{\alpha}_j^2\right)\right|\right].
\end{split}
\end{equation}
Here, we have defined $\bar{n}_j\equiv\bra{\phi_j}\hat{a}^\dagger\hat{a}\ket{\phi_j}$, $\bar{\alpha}_j\equiv\bra{\phi_j}\hat{a}\ket{\phi_j}$, and $\bar{\xi}_j=\bra{\phi_j}\hat{a}^2\ket{\phi_j}$. The minimization is over all possible decompositions (ensembles) $\{q_j,\ket{\phi_j}\}$ of $\hat{\rho}$, satisfying $\hat{\rho}=\sum_jq_j\ket{\phi_j}\bra{\phi_j}$ with $q_j>0$. If $\hat{\rho}$ is mixed (i.e., $\textrm{rank}(\hat{\rho})>1$), then the number of states in a decomposition of $\hat{\rho}$ may be any number larger than $\textrm{rank}(\hat{\rho})$, making the calculation of $\mathcal{N}(\hat{\rho})$ a highly complex optimization problem in general, hence this paper. 

The definition of $\mathcal{N}(\hat{\rho})$ satisfies several important conditions, as proved in Ref. \cite{ge2020operational}. (i) \textit{Non-negativity}: $\mathcal{N}(\hat{\rho})\geq0$, where equality holds if and only if $\hat{\rho}$ is classical. Non-negativity is the minimum requirement for a nonclassicality measure. (ii) \textit{Weak monotonicity}: $\mathcal{N}$ cannot increase under any classical operation $\Lambda$: $\mathcal{N}(\Lambda[\hat{\rho}])\leq\mathcal{N}(\hat{\rho})$. A classical operation is an operation that cannot create superpositions of coherent states from mixtures of coherent states (such as the use of passive linear optical operations and displacements). For details, see Ref. \cite{ge2020operational,ge2020evaluating}. Weak-monotonicity makes the ORT measure resource-theoretic; classical operations are the free operations of the resource theory. (iii) \textit{Convexity}: $\sum_jp_j\mathcal{N}(\hat{\rho}_j)\geq\mathcal{N}\left(\sum_jp_j\hat{\rho}_j\right)$ for any quantum states $\hat{\rho}_j$ and probabilities $p_j$. (iv) \textit{Lower-bounded by metrological power}: $\mathcal{N}(\hat{\rho})\geq\mathcal{W}(\hat{\rho})$, where equality holds for pure states. The metrological power for mixed states is defined as $\mathcal{W}(\hat{\rho})\equiv\max[F_X(\hat{\rho})-1/2,0]$, where \cite{ge2020operational,ge2020evaluating}
\begin{equation}\label{eq:FisherInformationMixed}
F_X(\hat{\rho})=\max_\mu\left[\min_{\{q_j,\ket{\phi_j}\}}\left(\sum_jq_j\bra{\phi_j}(\Delta \hat{X}_\mu)^2\ket{\phi_j}\right)\right]
\end{equation}
is the QFI for the optimal quadrature angle. Notice that for mixed states, the QFI also involves a minimization over all possible ensembles (and maximization over quadratures), although the order of the maximization and minimization in Eq. ~\eqref{eq:ORTMeasureMixed} and ~\eqref{eq:FisherInformationMixed} makes a subtle yet important difference (in fact, the QFI can be computed given an eigenstate decomposition of $\hat{\rho}$ \cite{toth2014quantum}). Again, the metrological power $\mathcal{W}$ describes the sensing enhancement over classical states: for all classical states, $\mathcal{W}=0$. Notably, there exist nonclassical mixed states for which $\mathcal{W}=0$ \cite{ge2020operational,ge2020evaluating}. Thus, the tight inequality $\mathcal{N}(\hat{\rho})\geq\mathcal{W}(\hat{\rho})$ is the closest possible relationship between a resource-theoretic nonclassicality measure and the metrological power.

\section{Method}

The ORT measure for mixed states, Eq. ~\eqref{eq:ORTMeasureMixed}, is an example of a \textit{convex roof construction}. Such constructions commonly arise in other nonclassicality measures~\cite{gehrke2012quantification, Kwon19, YadinPRX18}, as well as entanglement measures, for mixed states~\cite{toth2015evaluating, chen2005entanglement}. 

Evaluating the convex roof is a matter of optimizing an objective function over the set of all convex decompositions of a state $\hat{\rho}$. Optimizing the objective function is a challenging problem, a primary difficulty being that the number of states in a decomposition of a mixed state may be any number larger than its rank, although some clever techniques that bound or approximate the answer have been proposed \cite{toth2015evaluating,audenaert2001variational,rothlisberger2012libcreme}.

Rather than adapt previous techniques to our problem, here we introduce a method based on linear programming. Our method is similar to variational methods for ground state energy calculations, in that we essentially make a variational ansatz over the space of decompositions. It has the pedagogical advantage that it is highly intuitive and easy to employ. 

\subsection{Fock-diagonal states}
We utilize our method to evaluate the ORT measure for mixed states that are diagonal in the Fock basis: $\hat{\rho}=\sum_np_n\ket{n}\bra{n}$, where $\ket{n}$ is the Fock state with photon number $n$ and $\{p_n,\ket{n}\}$ is one possible decomposition. Such states arise often due to dephasing effects~\cite{Ferrini10,Arqand2020,Strzalka2024}, which eliminate the coherence between different Fock states. We refer to the probabilities $p_n$ as \textit{populations}. For such states, the ORT measure simplifies to:
\begin{equation}\label{eq:ORTFockDiagonal}
\mathcal{N}(\hat{\rho})=\langle{\hat{a}^\dagger\hat{a}}\rangle-\max_{\{q_j,\ket{\phi_j}\}}\left(\sum_j q_j\left|\Bar{\alpha}_j\right|^2-\left|\sum_j q_j\Bar{\alpha}_j^2\right|\right).
\end{equation}
It is obvious that $\mathcal{N}(\hat{\rho})\leq\langle\hat{a}^\dagger\hat{a}\rangle$, which is saturated by states in which all nonzero populations have associated photon numbers that differ by at least $2$ from one another \cite{ge2020evaluating}. An example is $p\ket{0}\bra{0}+(1-p)\ket{2}\bra{2}$, whose population constraints imply that any decomposition state has the form $\ket{\phi_j}=\ket{0}\bra{0}\ket{\phi_j}+\ket{2}\bra{2}\ket{\phi_j}$ --- it follows that $\alpha_j=\bra{\phi_j}\hat{a}\ket{\phi_j}=0$ for all $\ket{\phi_j}$. It is thus more worthwhile to evaluate the ORT measure for Fock-diagonal states where \textit{neighboring} photon numbers have nonzero populations.

Prior ORT measure calculations~\cite{ge2020evaluating} only addressed Fock-diagonal states with \textit{two} neighboring populations: $\hat{\rho}_{2F}=p_{n+1}\ket{n+1}\bra{n+1}+(1-p_{n+1})\ket{n}\bra{n}$. It was found that the optimal decomposition for such states was a ``four-prong" set of superposition states: $\ket{\phi_j}=\sqrt{p_{n+1}}\ket{n+1}+\sqrt{1-p_{n+1}}e^{ij\pi/2}\ket{n}$ with equal weights $q_j=1/4$ ($j=0,1,2,3$).\footnote{Actually, there are multiple optimal decompositions (for example a ``three-prong set" with relative phases that are the cubic roots of unity), but they share the property that $\left|\bra{n+1}\ket{\phi_j}\right|=\sqrt{p_{n+1}}$ for all $\ket{\phi_j}$.} The phases $e^{ij\pi/2}$, which are the quartic roots of unity, cause $\sum_jq_j\bar{\alpha}_j^2=0$ --- thus the absolute value term in Eq. ~\eqref{eq:ORTFockDiagonal} vanishes with this decomposition. Plugging this decomposition into Eq. ~\eqref{eq:ORTFockDiagonal} gives: $\mathcal{N}(\hat{\rho}_{2F})=n+p_{n+1}-(n+1)p_{n+1}(1-p_{n+1})$.\footnote{That this value is optimal may be further justified via its convexity as a function of $p_{n+1}$.} This result serves as a basic check of our method, and provides some insight into the types of optimal decompositions one may expect in more complicated cases.

\subsection{Linear-programming method}

Now we will describe our method for calculating the ORT measure. For simplicity, we will first explain it in the context of calculating $\mathcal{N}(\hat{\rho}_{2F})$. Then we will explain how to apply the method to Fock-diagonal states with an arbitrary number of nonzero populations.

Any state occurring in a decomposition of $\hat{\rho}$ must be in the support of $\hat{\rho}$. Thus, for $\hat{\rho}_{2F}$, each decomposition state must be of the form $\ket{\phi(x,\theta)}=x\ket{n+1}+\sqrt{1-x^2}e^{i\theta}\ket{n}$, where $x\in[0,1]$ and $\theta\in[0,2\pi]$. Just as there are infinitely many decompositions of $\hat{\rho}_{2F}$, there are infinitely many probability distributions $q(x,\theta)$ such that $\int_0^1dx\int_0^{2\pi}d\theta q(x,\theta)\ket{\phi(x,\theta)}\bra{\phi(x,\theta)}=\hat{\rho}_{2F}$. We prefer to think of the problem of finding the optimal decomposition as a problem of finding the optimal probability distribution $q(x,\theta)$. Each $q(x,\theta)$ satisfies the constraints:
\begin{subequations}\label{eq:constraintsqXTheta}
\begin{align}
\int_0^1dx\int_0^{2\pi}d\theta q(x,\theta)&=1 \label{eq:constraintsqXThetaNorm}\\
\int_0^1dx\int_0^{2\pi}d\theta x^2q(x,\theta)&=p_{n+1} \label{eq:constraintsqXThetaProb}\\
\int_0^1dx\int_0^{2\pi}d\theta x\sqrt{1-x^2}e^{i\theta}q(x,\theta)&=0 \label{eq:constraintsqXThetaCoherence}\\
q(x,\theta)&\geq0. \label{eq:constraintsqXThetaPositivity}
\end{align}
\end{subequations}
Respectively, the above constraints address normalization, population constraints, lack of nonzero off-diagonal elements, and positivity of all contributions (i.e., that the distribution describes a convex combination of states). Although we express $q(x,\theta)$ as if it were a continuous probability distribution here, the optimal distribution (decomposition) may be discrete. In terms of $q(x,\theta)$, Eq. ~\eqref{eq:ORTFockDiagonal} reads:
\begin{equation}\label{eq:ORTAsPDF}
\begin{split}
\mathcal{N}(\hat{\rho}_{2F})-\langle{\hat{a}^\dagger\hat{a}}\rangle=-\max_{q(x,\theta)}\Bigg(&\int_0^1dx \int_0^{2\pi}d\theta q(x,\theta)\left|\bar{\alpha}_{x,\theta}\right|^2\\
-&\left|\int_0^1dx \int_0^{2\pi}d\theta q(x,\theta)\bar{\alpha}_{x,\theta}^2\right|\Bigg)
\end{split}
\end{equation}
where $\bar{\alpha}_{x,\theta}=x\sqrt{1-x^2}\sqrt{n+1}e^{-i\theta}$. The quantity to be maximized on the right hand side of Eq.~\eqref{eq:ORTAsPDF} is called the objective function $\mathcal{L}[q(x,\theta)]$:
\begin{equation}\label{eq:objective}
\begin{split}
\mathcal{L}[q(x,\theta)]&=\int_0^1dx \int_0^{2\pi}d\theta q(x,\theta)\left|\bar{\alpha}_{x,\theta}\right|^2\\
&-\left|\int_0^1dx \int_0^{2\pi}d\theta q(x,\theta)\bar{\alpha}_{x,\theta}^2\right|.
\end{split}
\end{equation}
The second term in the objective function may appear problematic, due to the absolute value. However, without loss of generality, it can be taken to vanish (for our class of states). This is because, given any $q(x,\theta)$, it is always possible to find a distribution $\tilde{q}(x,\theta)$ such that $\mathcal{L}[\tilde{q}(x,\theta)]\geq\mathcal{L}[q(x,\theta)]$ and $\int_0^1dx \int_0^{2\pi}d\theta \tilde{q}(x,\theta)\bar{\alpha}_{x,\theta}^2=0$. The idea is to reallocate the probability from $q(x,\theta)$ in a four-prong formation for each $x$: $\tilde{q}(x,\theta)=\frac{1}{4}\left(\int_0^{2\pi}d\theta'q(x,\theta')\right)\left(\sum_{k=0}^{1}\sum_{j=0}^3\delta(\theta-\theta_0-\frac{j\pi}{2}-2\pi k)\right)$, where $\theta_0\in[0,\frac{\pi}{2})$ is arbitrary. It is straightforward to check that $\tilde{q}(x,\theta)$ satisfies the same constraints (Eq. ~\eqref{eq:constraintsqXThetaNorm}-~\eqref{eq:constraintsqXThetaPositivity}) as the $q(x,\theta)$ from which it was constructed; meanwhile, it causes the absolute term in the objective function $\mathcal{L}$ to vanish, while maintaining the value of the remaining term.

Since the $\theta$-dependence of $\tilde{q}(x,\theta)$ is solved, we may consider instead $\mathcal{Q}(x)\equiv\int_0^{2\pi} d\theta\tilde{q}(x,\theta)$. For our purposes, it is sufficient to optimize the $x$-dependence of $\mathcal{Q}(x)$, subject to the constraints $\int_0^1dx\mathcal{Q}(x)=1$, $\int_0^1dxx^2\mathcal{Q}(x)=p_{n+1}$, and $\mathcal{Q}(x)\geq0$. In terms of $\mathcal{Q}(x)$, Eq. ~\eqref{eq:ORTAsPDF} becomes:
\begin{equation}\label{eq:ORTAsPDF2}
\mathcal{N}(\hat{\rho}_{2F})-\langle{\hat{a}^\dagger\hat{a}}\rangle
=-\max_{\mathcal{Q}(x)}\left(\int_0^1dx \mathcal{Q}(x)x^2(1-x^2)(n+1)\right).
\end{equation}
The optimization problem (including constraints) resembles a \textit{linear programming} problem \cite{boyd2004convex} for the vector $\mathcal{Q}(x)$, the only caveat being that $\mathcal{Q}(x)$ does not have a discrete number of components. Nevertheless, the solution to the problem can be approximated by discretizing the set of possible $x$ (i.e., binning):
\begin{equation}\label{eq:ORTDiscrete2F}
\begin{split}
-\max_{\mathcal{Q}_x} \left(\sum_{x\in \mathcal{X}} \mathcal{Q}_x x^2(1-x^2)(n+1)\right)
&\gtrsim\mathcal{N}(\hat{\rho}_{2F})-\langle{\hat{a}^\dagger\hat{a}}\rangle\\
\end{split}
\end{equation}
\begin{subequations}\label{eq:constraintsQxTwoFock}
\begin{align}
\sum_{x\in \mathcal{X}}\mathcal{Q}_x&=1\\
\sum_{x\in \mathcal{X}}x^2\mathcal{Q}_x&=p_{n+1}\\
\mathcal{Q}_x&\geq0.
\end{align}
\end{subequations}
Here, $\mathcal{X}$ is a discrete set of points in the range $[0,1]$. In this work, we take $\mathcal{X}$ to be the set of real numbers $m\Delta\leq 1$, where $m$ is a non-negative integer and $\Delta$ is the spacing of the binning with $0<\Delta\ll1$). The approximation, Eq. ~\eqref{eq:ORTDiscrete2F}, should converge as $\Delta\rightarrow0$ (here $\gtrsim$ means approximately equal to, but never less than).
\begin{figure}
    \centering
    \includegraphics[width=0.45 \textwidth]{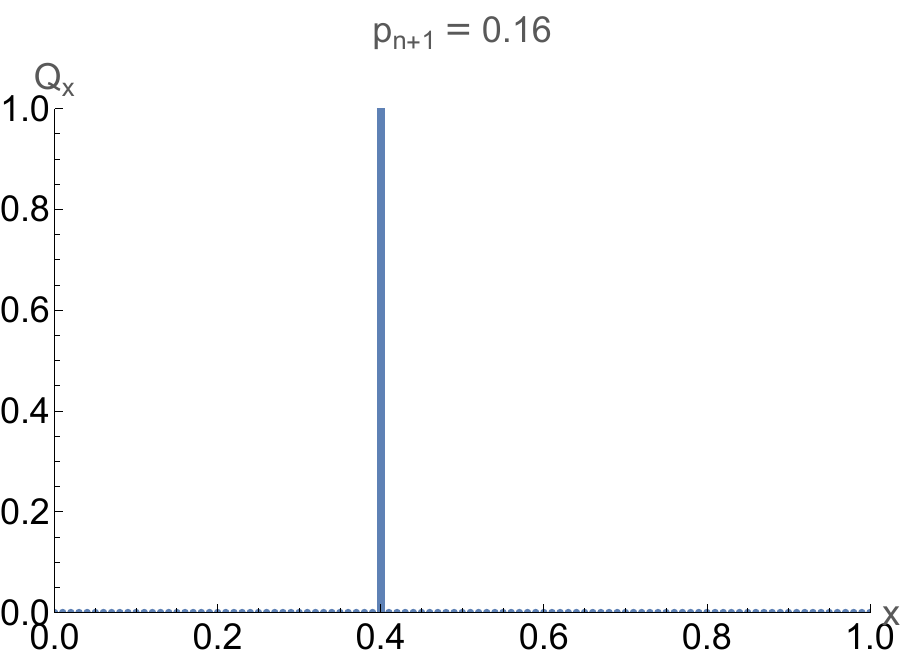}
    \caption{Optimal discrete probability distribution $\mathcal{Q}_x$ for $\hat{\rho}_{2F}=0.16\ket{n+1}\bra{n+1}+0.84\ket{n}\bra{n}$ (the value of $n$ has no effect here). The spacing between $x$ bins is $\Delta=0.01$. One may notice that $\mathcal{Q}_{\sqrt{p_{n+1}}}=1$ and $\mathcal{Q}_x=0$ for all other $x$, thus reproducing the expected result from Ref. \cite{ge2020evaluating}.}
    \label{fig:rho2F}
\end{figure}

Here $\mathcal{Q}_x$ is a discrete probability distribution (``histogram") over the set of points in $\mathcal{X}$, satisfying constraints which reproduce the original density matrix. As such, $\mathcal{Q}_x$ is essentially a variational ansatz for the optimal decomposition, which can be optimized by linear programming. We solve the linear program using the \textsc{LinearOptimization} function in \textsc{Mathematica}. An example is demonstrated in Fig. \ref{fig:rho2F}. $\mathcal{Q}_x$ is spiked at $x=\sqrt{p_{n+1}}$ and zero elsewhere. This holds for all choices of $p_{n+1}$ (though one should be mindful and bin appropriately in the extremes $p_{n+1}\ll1$ and $1-p_{n+1}\ll1$). Thus, the optimal decomposition utilizes the four-prong set of states $\ket{\phi_j}=\sqrt{p_{n+1}}\ket{n+1}+\sqrt{1-p_{n+1}}e^{i(j\pi/2+\theta_0)}$ (where $j=0,1,2,3$), with probability $1/4$ for each. This corroborates our method, since it reproduces the results of Ref. \cite{ge2020evaluating}.

Note that, due to a careful choice of parameters (and binning) for Fig. \ref{fig:rho2F}, the discrete distribution from linear optimization matches the true optimal decomposition perfectly. This will not always be the case, since the chosen binning may skip over the optimal point(s) (generally, the discrete distribution will be slightly suboptimal, acting as a bound). However, one can work backwards from the numerical evidence to make an ansatz for a better decomposition (which may, in fact, be the true optimal one). For example, by realizing that $0.4=\sqrt{0.16}$, in Fig. \ref{fig:rho2F}, one may conjecture the general trend that $\mathcal{Q}_x$ is spiked at $x=\sqrt{p_{n+1}}$.  Such an ansatz can then be corroborated by checking it against finer binnings and other values for the populations. In this way, we obtain analytical ansatzes for $\mathcal{N}(\hat{\rho}_{3F})$ and $\mathcal{N}(\hat{\rho}_{4F})$, in section \ref{sec:results}.

\subsection{Higher-rank Fock-diagonal states}\label{subsec:higherRankFock}
A more general Fock-diagonal state, with neighboring populations, is:
\begin{equation}\label{eq:MFock}
\hat{\rho}_{MF}=\sum_{k=0}^{M-1}p_{n+k}\ket{n+k}\bra{n+k}.
\end{equation}
Any decomposition state for $\hat{\rho}_{MF}$ must be of the form:
\begin{equation}\label{eq:genericDecompositionState}
\ket{\phi\left(\vec{x},\vec{\theta}\right)}=\sum_{k=0}^{M-1}x_ke^{i\theta_k}\ket{n+k},
\end{equation}
where $x_k\in[0,1]$, $\theta_k\in[0,2\pi]$, and $\sum_k x_k^2=1$. Without loss of generality, one can take $\theta_{M-1}=0$, and $x_0=\sqrt{1-\sum_{k=1}^{M-1}x_k^2}$. $\hat{\rho}_{MF}$ may then be expressed as a mixture of these states with probabilities $q(\vec{x},\vec{\theta})$ satisfying certain constraints. The nonclassicality calculation then becomes a matter of optimizing $q(\vec{x},\vec{\theta})$. Similar to the case of $\hat{\rho}_{2F}$, it is sufficient to optimize a probability distribution $\mathcal{Q}(\vec{x})\equiv\sum_{\vec{\theta}}\tilde{q}(\vec{x},\vec{\theta})$ over the vectors $\vec{x}$. This is because, given any $q(\vec{x},\vec{\theta})$, one may always construct (see Appendix \ref{appendixB}) a $\tilde{q}(\vec{x},\vec{\theta})$ with $\vec{\theta}$-dependence such that:
\begin{equation}\label{eq:tildeQHigherRank}
\frac{1}{\sum_{\vec{\theta'}}q(\vec{x},\vec{\theta'})}\sum_{\vec{\theta}}\tilde{q}(\vec{x},\vec{\theta})\ket{\phi\left(\vec{x},\vec{\theta}\right)}\bra{\phi\left(\vec{x},\vec{\theta}\right)}=\hat{\rho}_{\vec{x}},
\end{equation}
where $\hat{\rho}_{\vec{x}}=\sum_{k=0}^{M-1}x_k^2\ket{n+k}\bra{n+k}$ is a normalized Fock-diagonal state indexed by $\vec{x}$, $\sum_{\vec{\theta}}\tilde{q}(\vec{x},\vec{\theta})|\bar{\alpha}_{\vec{x},\vec{\theta}}|^2\geq\sum_{\vec{\theta}}q(\vec{x},\vec{\theta})|\bar{\alpha}_{\vec{x},\vec{\theta}}|^2$, and $\sum_{\vec{\theta}}\tilde{q}(\vec{x},\vec{\theta})\bar{\alpha}_{\vec{x},\vec{\theta}}^2=0$. Constructed in this way, $\tilde{q}(\vec{x},\vec{\theta})$ will be at least as optimal as the originating $q(\vec{x},\vec{\theta})$. Note that the existence of such a $\tilde{q}(\vec{x},\vec{\theta})$ demonstrates that Eq. ~\eqref{eq:ORTFockDiagonal} may be rewritten:
\begin{equation}\label{eq:ORTFockDiagonalRewrite}
\mathcal{N}(\hat{\rho}_{MF})=\langle{\hat{a}^\dagger\hat{a}}\rangle-\max_{\{q_j,\ket{\phi_j}\}}\left(\sum_j q_j\left|\Bar{\alpha}_j\right|^2\right),
\end{equation}
in general, for Fock-diagonal states.

The optimization problem for $\mathcal{Q}(\vec{x})$ can be solved approximately by discretizing the set of possible $\vec{x}$:
\begin{equation}\label{eq:ORTDiscreteMF}
\begin{split}
-\max_{\mathcal{Q}_{\vec{x}}}\left[\sum_{\vec{x}\in\mathcal{X}}\mathcal{Q}_{\vec{x}}\left(\sum_{k=0}^{M-2}x_{k+1}x_k\sqrt{n+k+1}\right)^2\right]\\
\gtrsim \mathcal{N}(\hat{\rho}_{MF})-\langle\hat{a}^\dagger\hat{a}\rangle
\end{split}
\end{equation}
\begin{subequations}\label{eq:constraintsQxDiscrete}
\begin{align}
\sum_{\vec{x}\in\mathcal{X}}\mathcal{Q}_{\vec{x}}&=1 \label{eq:QxNormalization}\\
\sum_{\vec{x}\in\mathcal{X}}x_k^2\mathcal{Q}_{\vec{x}}&=p_{n+k} \label{eq:QxPopulation}\\
\mathcal{Q}_{\vec{x}}&\geq0.
\end{align}
\end{subequations}
In this work, we take $\left(x_1,x_2,...,x_{M-1}\right)=\left(l_1\Delta,l_2\Delta,...,l_{M-1}\Delta\right)$, where $\sum_{k=1}^{M-1}l_k^2\Delta^2\leq1$, $l_k$ are non-negative integers, and $0<\Delta\ll1$. The approximation, Eq. ~\eqref{eq:ORTDiscreteMF}, should converge as $\Delta\rightarrow0$. Note that $\mathcal{Q}_{\vec{x}}$ is essentially a vector in $\mathbb{R}^d$, where $d$ is the number of elements in $\mathcal{X}$. For fixed $\Delta$, $d$ scales as $\Delta^{-M+1}$. The optimization may be carried out using linear programming, as before, although the computational complexity will increase with $M$. 

\subsection{Simply- and Compositely-decomposed States}\label{subsec:simplyAndCompositelyDecomposedStates}

Note that an upper bound on the ORT measure for Fock-diagonal states can easily be found by plugging in the simple decomposition $\mathcal{Q}_{\vec{x}}=\prod_{k=1}^{M-1}\delta_{x_k,\sqrt{p_k}}$:
\begin{equation}\label{eq:simpleBound}
\mathcal{N}(\hat{\rho}_{MF})\leq\langle\hat{a}^\dagger\hat{a}\rangle-\left(\sum_{k=0}^{M-2}\sqrt{p_{n+k+1}p_{n+k}(n+k+1)}\right)^2.
\end{equation}
Equality holds in the $M=2$ case (where this simple decomposition is optimal), as demonstrated previously. For $M>2$, equality holds in some regimes (i.e., some choices of the populations $p_{n+k}$), although not all.

States that saturate the equality of Eq. ~\eqref{eq:simpleBound} are of some fundamental interest, acting as building blocks to decompose other states. As implied by Eq. ~\eqref{eq:tildeQHigherRank} and ~\eqref{eq:QxNormalization}-~\eqref{eq:QxPopulation}, for any probability distribution $\mathcal{Q}(\vec{x})$: 
\begin{equation}\label{eq:compositeDecomposition}
\sum_{\vec{x}}\mathcal{Q}(\vec{x})\hat{\rho}_{\vec{x}}=\hat{\rho}_{MF}.
\end{equation}
We claim that, for the optimal $\tilde{\mathcal{Q}}(\vec{x})$, any $\hat{\rho}_{\vec{x}}$ (which is itself a Fock-diagonal state) for which $\tilde{\mathcal{Q}}(\vec{x})>0$ must saturate the equality of Eq. ~\eqref{eq:simpleBound}. To see this, suppose that for some contributing $\vec{x}^*$, $\hat{\rho}_{\vec{x}^*}$ does not saturate Eq. ~\eqref{eq:simpleBound}. This would imply that there exists an optimal decomposition $\tilde{\mathcal{R}}(\vec{y})$ of $\hat{\rho}_{\vec{x}^*}$ such that: $\sum_{\vec{y}}\tilde{\mathcal{R}}({\vec{y}})\left(\sum_{k=0}^{M-2}y_{k+1}y_k\sqrt{n+k+1}\right)^2$ is greater than $\left(\sum_{k=0}^{M-2}x_{k+1}^*x_k^*\sqrt{n+k+1}\right)^2.$ One could then construct a better decomposition (than $\tilde{\mathcal{Q}}(\vec{x})$) of $\hat{\rho}_{MF}$ by reallocating the probability assigned to $\hat{\rho}_{\vec{x}^*}$: $\tilde{\mathcal{S}}(\vec{x})=\tilde{\mathcal{Q}}(\vec{x})(1-\delta_{\vec{x},\vec{x}^*})+\tilde{\mathcal{Q}}(\vec{x}^*)\tilde{\mathcal{R}}(\vec{x})$. But, by assumption, $\mathcal{Q}(\vec{x})$ was optimal. Thus, there is a contradiction with having $\hat{\rho}_{\vec{x}^*}$ not saturate Eq. ~\eqref{eq:simpleBound}. Our results reflect this, as discussed in subsection \ref{subsec:remarks}.

We call Fock-diagonal states that saturate the equality of Eq. ~\eqref{eq:simpleBound} \textit{simply-decomposed} states, and Fock-diagonal states that do not saturate Eq. ~\eqref{eq:simpleBound} \textit{compositely-decomposed} states. The optimal decomposition of a compositely decomposed state is a convex combination $\hat{\rho}_{MF}=\sum_{\vec{x}}\tilde{\mathcal{Q}}(\vec{x})\hat{\rho}_{\vec{x}}$ of simply-decomposed states. However, each contributing $\hat{\rho}_{\vec{x}}$ could have a rank less than or equal to $M$.

\section{Numerical Results and Analytical ansatz}\label{sec:results}

\subsection{Rank-3 case}
\begin{figure}
    \centering
    \includegraphics[width=0.45\textwidth]{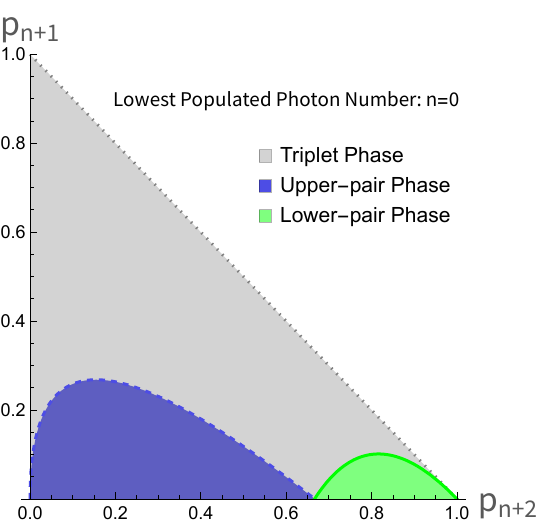}
    \caption{Phase diagram for $\hat{\rho}_{3F}$ assuming the lowest populated photon number is $n=0$. The expression for the ORT measure $\mathcal{N}(\hat{\rho}_{3F})$ depends on which of three distinct phases the state is in (although $\mathcal{N}(\hat{\rho}_{3F})$ is still piecewise continuous as shown in Fig.~\ref{fig:fullORT3}). These phases are named according to what their optimal decompositions look like. For $n>0$, the general shape of these phases is maintained, although the point where the curves intersect the bottom axis shifts left as $p_{n+2}=\frac{2+n}{3+2n}$.}
    \label{fig:rank3PhaseSpace}
\end{figure}

For fixed $n$ in $\hat{\rho}_{3F}=p_{n+2}\ket{n+2}\bra{n+2}+p_{n+1}\ket{n+1}\bra{n+1}+(1-p_{n+2}-p_{n+1})\ket{n}\bra{n}$, there are two free parameters: $p_{n+2}$ and $p_{n+1}$. The possible choices $(p_{n+2},p_{n+1})$ of these parameters form the 45-45-90 triangle defined by $p_{n+2}+p_{n+1}\leq1$, $p_{n+2}\geq0$, and $p_{n+1}\geq0$. From our numerical investigations, we identified three distinct regimes, or \textit{phases}, within this triangular phase space (see Fig. \ref{fig:rank3PhaseSpace}). For each phase, we provide an analytical ansatz, based on numerical evidence, for the ORT measure and associated optimal decomposition.

\subsubsection{Triplet phase}
In Fig. \ref{fig:rank3PhaseSpace}, the triplet phase is the upper, gray regime (the lower boundary of which is given in ensuing subsections). Our ansatz for the optimal decomposition of $\hat{\rho}_{3F}$ uses states of the form:
\begin{equation}\label{eq:phiOptimalTriplet}
\begin{split}
\ket{\phi_T(\theta_j,\gamma_j)}&=\sqrt{p_{n+2}}\ket{n+2}+\sqrt{p_{n+1}}e^{i\theta_j}\ket{n+1}\\
&+\sqrt{1-p_{n+2}-p_{n+1}}e^{i\gamma_j}\ket{n}.
\end{split}
\end{equation}
For more information about the probabilities assigned to these states, and their phases, see Appendix \ref{appendixB}. For context, this decomposition is a straightforward generalization of the optimal decomposition for density matrices that are diagonal in \textit{two} neighboring Fock states, $\hat\rho_{2F}=p_{n+1}\ket{n+1}\bra{n+1}+p_n\ket{n}\bra{n}$ (see section III and \cite{ge2020evaluating}). The ORT measure for states in this regime is:
\begin{equation}\label{eq:ORTMeasureTriplet}
\begin{split}
&\mathcal{N}(\hat{\rho}_{3F})\lesssim2p_{n+2}+p_{n+1}+n-\\
&\left(\sqrt{p_{n+2}}\sqrt{n+2}+\sqrt{1-p_{n+2}-p_{n+1}}\sqrt{n+1}\right)^2p_{n+1}.
\end{split}
\end{equation}
We conjecture (without proof) that the inequality is saturated throughout the triplet phase, hence the ``$\lesssim$" symbol. Assuming this is correct, the bound in Eq. ~\eqref{eq:simpleBound} is saturated.

What stands out about this decomposition is the fact that $\left|\braket{n+2}{\phi_T(\theta_j,\gamma_j)}\right|^2=p_{n+2}$, $\left|\bra{n+1}\ket{\phi_T(\theta_j,\gamma_j)}\right|^2=p_{n+1}$, and $\left|\bra{n}\ket{\phi_T(\theta_j,\gamma_j)}\right|^2=1-p_{n+2}-p_{n+1}$. Therefore, we call this decomposition the ``triplet" decomposition (and this phase the triplet phase).

Of course, it is always possible to decompose $\hat{\rho}_{3F}=p_{n+2}\ket{n+2}\bra{n+2}+p_{n+1}\ket{n+1}\bra{n+1}+p_{n}\ket{n}\bra{n}$ into the four states $\ket{\phi_T(\theta_j,\gamma_j)}$. However, it turns out that the triplet decomposition is \textit{not} optimal for all choices of $(p_{n+2},p_{n+1})$. For some values of $(p_{n+2},p_{n+1})$, a more optimal decomposition is possible.

\subsubsection{Upper-pair phase}
In Fig. \ref{fig:rank3PhaseSpace}, the upper-pair phase is the lower-left, blue regime. Our ansatz for the optimal decomposition of $\hat{\rho}_{3F}$ uses the Fock state $\ket{n}$ and states of the form:
\begin{small}
\begin{equation}\label{eq:phiOptimalUpperPair}
\begin{split}
&\ket{\phi_U(\theta_j,\gamma_j)}=\sqrt{1-f}e^{i\gamma_j}\ket{n}\\
&+\sqrt{f}\left(\sqrt{\frac{p_{n+2}}{p_{n+2}+p_{n+1}}}\ket{n+2}+\sqrt{\frac{p_{n+1}}{p_{n+2}+p_{n+1}}}e^{i\theta_j}\ket{n+1}\right),
\end{split}
\end{equation}
\end{small}
where:
\begin{equation}\label{eq:optimalF}
f=\frac{(2+n)p_{n+2}}{(1+n)p_{n+1}+(3+2n)p_{n+2}}.
\end{equation}
The states $\ket{\phi_U(\theta_j,\gamma_j)}$ contribute a total probability $\left(p_{n+2}+p_{n+1}\right)f^{-1}$. The ORT measure is:
\begin{equation}\label{eq:ORTMeasureUpperPair}
\begin{split}
\mathcal{N}(\hat{\rho}_{3F})&\lesssim2p_{n+2}+p_{n+1}+n\\
&-(p_{n+2}+p_{n+1})f^{-1}\left|\bra{\phi_U(0,0)}\hat{a}\ket{\phi_U(0,0)}\right|^2.
\end{split}
\end{equation}

What stands out about this decomposition is the fact that $\left|\bra{n+2}\ket{\phi_U(\theta_j,\gamma_j)}\right|^2/\left|\bra{n+1}\ket{\phi_U(\theta_j,\gamma_j)}\right|^2=p_{n+2}/p_{n+1}$, the population ratio between $\ket{n+2}$ and $\ket{n+1}$ in the original density matrix $\hat{\rho}_{3F}$. By contrast, $\left|\bra{n}\ket{\phi_U(\theta_j,\gamma_j)}\right|^2$ does not have a neat connection to the population ratios in the original density matrix. Moreover, $\ket{n}$ is also a state in the decomposition. Due to this asymmetric treatment of the photon numbers, we call this the upper-pair phase.

One may wonder how the value of $f$ was obtained, given its unintuitive appearance. First, we conjectured based on the linear optimization that, for some values of $(p_{n+2},p_{n+1})$, the optimal decomposition involved the Fock state $\ket{n}$ and states of the form $\ket{\phi_U(\theta_j,\gamma_j)}$ (see Eq. ~\eqref{eq:phiOptimalUpperPair}). We then optimized the value of $f$ in Eq. ~\eqref{eq:phiOptimalUpperPair} analytically, yielding Eq. ~\eqref{eq:optimalF}. This optimization is detailed in Appendix \ref{appendixA}.

Such a decomposition is not possible everywhere, though. As previously mentioned , the states $\ket{\phi_U(\theta_j,\gamma_j)}$ contribute a total probability $(p_{n+2}+p_{n+1})f^{-1}$, which cannot exceed $1$. Thus:
\begin{equation}\label{eq:upperPairBound}
p_{n+2}+p_{n+1}\leq\frac{(2+n)p_{n+2}}{(1+n)p_{n+1}+(3+2n)p_{n+2}}.
\end{equation}
This sets the upper bound on this phase (blue dashed curve in Fig. \ref{fig:rank3PhaseSpace}): note that, at the upper bound, $\ket{\phi_U(\theta_j,\gamma_j}=\ket{\phi_T(\theta_j,\gamma_j)}$. To find the highest possible value of $p_{n+2}$, we can set $p_{n+1}=0$ and saturate the inequality:
\begin{equation}\label{eq:pPlus2Max}
p_{n+2}^{\text{max}}=\frac{2+n}{3+2n}.
\end{equation}

\begin{figure}
    \centering
    \includegraphics[width=0.45\textwidth]{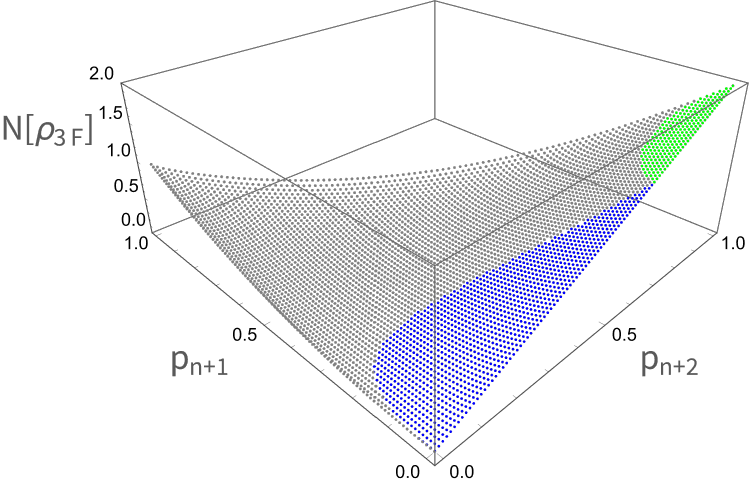}
    \caption{(Color online) Nonclassicality $\mathcal{N}(\hat{\rho}_{3F})$ over the entire phase space of $\hat{\rho}_{3F}$. Here, the lowest populated photon number is assumed to be $n=0$. At the corners, which represent Fock states, the nonclassicality takes the values $0$, $1$, and $2$, respectively. The nonclassicality is continuous despite there being three distinct phases (see Fig. \ref{fig:rank3PhaseSpace}). The points in each phase are colored differently here for reference.}
    \label{fig:fullORT3}
\end{figure}

\subsubsection{Lower-pair phase}

In Fig. \ref{fig:rank3PhaseSpace}, the lower-pair phase is the lower-right, green regime. It is analogous to the upper-pair phase, except that it is the lower-two Fock states $\ket{n+1}$ and $\ket{n}$ that are ``paired." Our ansatz for the optimal decomposition of $\hat{\rho}_{3F}$ uses the Fock state $\ket{n+2}$ and states of the form:
\begin{small}
\begin{equation}\label{eq:phiOptimalLowerPair}
\begin{split}
&\ket{\phi_L(\theta_j,\gamma_j)}=\sqrt{1-g}\ket{n+2}\\
&+\sqrt{g}\left(\sqrt{\frac{p_{n+1}}{1-p_{n+2}}}e^{i\theta_j}\ket{n+1}+\sqrt{\frac{1-p_{n+2}-p_{n+1}}{1-p_{n+2}}}e^{i\gamma_j}\ket{n}\right),
\end{split}
\end{equation}
\end{small}
where:
\begin{equation}\label{eq:optimalG}
g=\frac{(1+n)(-1+p_{n+1}+p_{n+2})}{-3-2n+(1+n)p_{n+1}+(3+2n)p_{n+2}}.
\end{equation}
The states $\ket{\phi_L(\theta_j,\gamma_j)}$ contribute a total probability $(1-p_{n+2})g^{-1}$. The ORT measure is:
\begin{equation}\label{eq:ORTMeasureLowerPair}
\begin{split}
\mathcal{N}(\hat{\rho}_{3F})&\lesssim2p_{n+2}+p_{n+1}+n\\
&-(1-p_{n+2})g^{-1}\left|\bra{\phi_L(0,0)}\hat{a}\ket{\phi_L(0,0)}\right|^2.
\end{split}
\end{equation}

We call this the ``lower-pair phase" due to the asymmetric treatment of the photon numbers. $\left|\bra{n+1}\ket{\phi_L(\theta_j,\gamma_j)}\right|^2/\left|\bra{n}\ket{\phi_L(\theta_j,\gamma_j)}\right|^2=p_{n+1}/(1-p_{n+2}-p_{n+1})$, the population ratio between $\ket{n+1}$ and $\ket{n}$ in the original state. By contrast, $\left|\bra{n+2}\ket{\phi_L(\theta_j,\gamma_j)}\right|^2$ does not have a simple connection to the population ratios in the original density matrix; also, $\ket{n+2}$ is itself a state in the decomposition.

The value of $g$ was obtained in the same way that $f$ in Eq. ~\eqref{eq:optimalF} was. We conjectured based on the linear optimization that, for some values of $(p_{n+2},p_{n+1})$, the optimal decomposition involved the Fock state $\ket{n+2}$ and states of the form $\ket{\phi_L(\theta_j,\gamma_j)}$ (see Eq. ~\eqref{eq:phiOptimalLowerPair}). We then optimized $g$ in Eq. ~\eqref{eq:phiOptimalLowerPair} analytically. See Appendix \ref{appendixA} for further details.

Similarly, the total probability $(1-p_{n+2})g^{-1}$ conttibuted by the states $\ket{\phi_L(\theta_j,\gamma_j)}$ cannot exceed $1$ such that 
\begin{equation}\label{eq:lowerPairBound}
1-p_{n+2}\leq\frac{(1+n)(-1+p_{n+1}+p_{n+2})}{-3-2n+(1+n)p_{n+1}+(3+2n)p_{n+2}}.
\end{equation}
This sets the upper bound on this phase (green solid curve in Fig. \ref{fig:rank3PhaseSpace}): note that, at the upper bound, $\ket{\phi_L(\theta_j,\gamma_j}=\ket{\phi_T(\theta_j,\gamma_j)}$. To find the lowest possible value of $p_{n+2}$, we can set $p_{n+1}=0$ and saturate the inequality:
\begin{equation}\label{eq:pPlus2Min}
p_{n+2}^{\text{min}}=\frac{2+n}{3+2n}.
\end{equation}

Having scanned over various values of $(p_{n+2},p_{n+1})$, and numerically approximated $\mathcal{N}(\hat{\rho}_{3F})$ we suspect (although we lack analytical proof) that these three phases (triplet, upper-pair, and lower-pair) are the only three, and that our ansatzes characterize the nonclassicality exactly over the entire phase space of $\hat{\rho}_{3F}$. Assuming our ansatzes, the nonclassicality is plotted in Fig. \ref{fig:fullORT3}

\subsubsection{Remarks}\label{subsec:remarks}
In subsection \ref{subsec:simplyAndCompositelyDecomposedStates}, we distinguished between simply-decomposed states, which saturate Eq. ~\eqref{eq:simpleBound}, and compositely-decomposed states, whose optimal decompositions use simply-decomposed states as building blocks. Our results reflect this. States in the triplet phase are simply-decomposed, while states in the upper-pair and lower-pair phase are compositely decomposed. For example, $\hat{\rho}_{\text{comp}}=0.2\ket{2}\bra{2}+0.2\ket{1}\bra{1}+0.6\ket{0}\bra{0}$ is compositely-decomposed, since it lies in the upper-pair phase (Fig. \ref{fig:rank3PhaseSpace}). Nevertheless, $\hat{\rho}_{\text{comp}}$ may be rewritten as a convex combination of density matrices that \textit{are} simply-decomposed: $\hat{\rho}_{\text{comp.}}=p_1\hat{\rho}_{\text{simp},1}+p_2\hat{\rho}_{\text{simp},2}$. In this case $\hat{\rho}_{\text{simp},1}=\ket{0}\bra{0}$,  $\hat{\rho}_{\text{simp},2}=0.25\ket{2}\bra{2}+0.25\ket{1}\bra{1}+0.5\ket{0}\bra{0}$, $p_1=0.2$, and $p_2=0.8$ --- interestingly, $\hat{\rho}_{\text{simp}, 2}$ is precisely at the boundary of the triplet and upper-pair phases (Fig. \ref{fig:rank3PhaseSpace}). The upper-pair decomposition of $\hat{\rho}_{\text{comp.}}$ makes precise use of this convex combination, see Eq. ~\eqref{eq:phiOptimalUpperPair} and ~\eqref{eq:optimalF}.

\subsection{Rank-4 case}
For fixed $n$ in $\hat{\rho}_{4F}=p_{n+3}\ket{n+3}\bra{n+3}+p_{n+2}\ket{n+2}\bra{n+2}+p_{n+1}\ket{n+1}\bra{n+1}+(1-p_{n+3}-p_{n+2}-p_{n+1})\ket{n}\bra{n}$, there are three free parameters: $p_{n+3}$, $p_{n+2}$, and $p_{n+1}$. The possible choices $(p_{n+3},p_{n+2},p_{n+1})$ of these parameters form the triangular pyramid defined by $p_{n+3}+p_{n+2}+p_{n+1}\leq1$, $p_{n+3}\geq0$, $p_{n+2}\geq0$, and $p_{n+1}\geq0$. From our numerical investigations, we identified six distinct regimes, or phases, within this pyramidal phase space. These fall into three categories: which we call \textit{quartet}, \textit{triplet}, and \textit{pair}. For each phase, we provide an ansatz for the ORT measure and associated optimal decomposition. The reader should regard the ansatzes for the ORT measure here as close and useful upper bounds, and these phases as a useful guide, akin to a cartographer's initial map of some region. Since identifying these regions, we \textit{have} discovered examples wherein the nonclassicality surpasses the bounds presented here, albeit by small margins (examples are provided in subsection \ref{subsec:exceptions}).

\begin{widetext}
\begin{figure*}
    \centering
    \includegraphics[width=\linewidth]{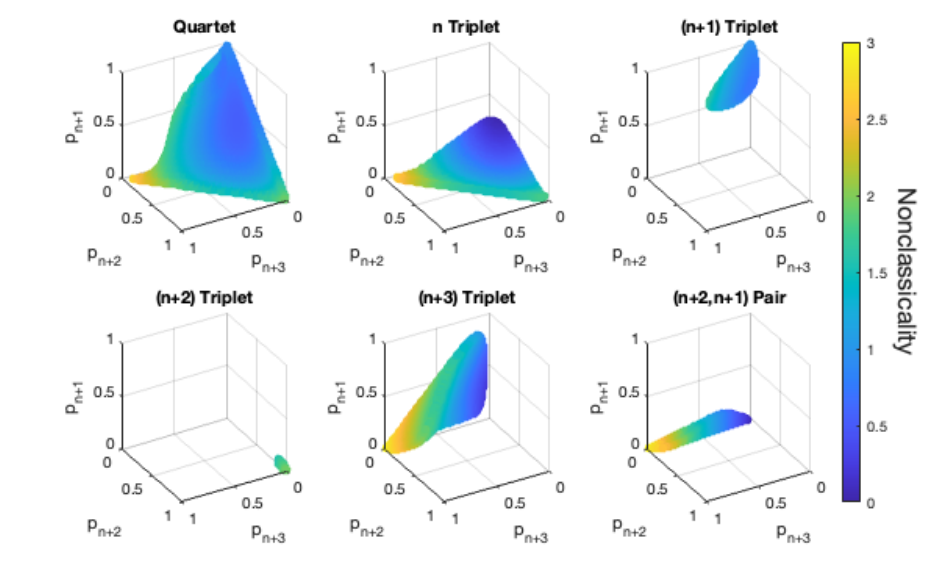}
    \caption{Nonclassicality $\mathcal{N}(\hat{\rho}_{4F})$ over the phase space of $\hat{\rho}_{4F}$, assuming the lowest populated photon number of $n=0$. There are six distinct regimes (phases) of optimally decomposing $\hat{\rho}_{4F}$. The complete phase space composed of these regimes is the triangular pyramid defined by $p_{n+3}+p_{n+2}+p_{n+1}\leq1$, $p_{n+3}\ge0$, $p_{n+2}\ge0$, and $p_{n+1}\ge0$. This plot was created using our analytical ansatzes at discrete points within each phase, and some phases may be larger than they appear. At the vertices of the pyramid (which represent Fock states), the ORT measure takes the values $0$, $1$, $2$, and $3$, respectively.}
    \label{fig:sixPhases}
\end{figure*}
\end{widetext}

\subsubsection{Quartet Phase}
The quartet phase is shown in Fig. \ref{fig:sixPhases} (upper-left). This phase comprises the majority of the phase space, in the same way that the triplet phase comprises the majority of the phase space in the rank-3 case. Our ansatz for the optimal decomposition here is the most simple, utilizing states of the form:
\begin{equation}\label{eq:phiOptimalQuartet}
\begin{split}
\ket{\phi_Q(\theta_j,\gamma_j,\zeta_j)}&=\sqrt{p_{n+3}}\ket{n+3}+\sqrt{p_{n+2}}e^{i\theta_j}\ket{n+2}\\
&+\sqrt{p_{n+1}}e^{i\gamma_j}\ket{n+1}+\sqrt{p_n}e^{i\zeta_j}\ket{n},
\end{split}
\end{equation}
where $p_n=1-\sum_{k=1}^3p_{n+k}$. These states are a straightforward generalization of the states in Eq. ~\eqref{eq:phiOptimalTriplet}. For more information about the probabilities assigned to these states, and the phase angles $\theta_j,\gamma_j,\xi_j$, see Appendix \ref{appendixB}. We call this the quartet phase because all four photon numbers are treated symmetrically, with amplitudes equal to the square root of the populations. The ORT measure for states in this regime is:
\begin{equation}\label{eq:ORTMeasureQuartet}
\begin{split}
\mathcal{N}(\hat{\rho}_{4F})&\lesssim3p_{n+3}+2p_{n+2}+p_{n+1}+n\\
&-\left(\sum_{k=0}^2\sqrt{p_{n+k+1}p_{n+k}(n+k+1)}\right)^2.
\end{split}
\end{equation}
Assuming equality, this saturates the bound in Eq. ~\eqref{eq:simpleBound}.

It is always possible to decompose $\hat{\rho}_{4F}$ into the states in Eq. ~\eqref{eq:phiOptimalQuartet}. However, we determined that this decomposition is not optimal for some choices of $(p_{n+3},p_{n+2},p_{n+1})$. Regions where better decompositions exist (and, consequently, the boundaries of the quartet phase) are described in the following.

\subsubsection{Triplet Phases}
The triplet phases are natural generalizations of the pair phases in the rank-3 case. There are four triplet phases: the $n$-triplet, $(n+1)$-triplet, $(n+2)$-triplet, and $(n+3)$-triplet.

\textit{$n$-triplet}---The $n$-triplet phase is shown in the upper-middle panel of Fig. \ref{fig:sixPhases}. Our ansatz for the optimal decomposition consists of one Fock state, $\ket{n}$, as well as states of the form:
\begin{small}
\begin{equation}\label{eq:phiOptimalnTriplet}
\begin{split}
&\ket{\phi_{0T}(\theta_j,\gamma_j,\zeta_j)}=\sqrt{1-f_0}e^{i\zeta_j}\ket{n}+\sqrt{f_0}\Bigg(\sqrt{\frac{p_{n+3}}{1-p_n}}\ket{n+3}\\
&+\sqrt{\frac{p_{n+2}}{1-p_n}}e^{i\theta_j}\ket{n+2}+\sqrt{\frac{p_{n+1}}{1-p_n}}e^{i\gamma_j}\ket{n+1}\Bigg).
\end{split}
\end{equation}
\end{small}
Such states contribute a total probability $(1-p_n)f_0^{-1}$, which cannot exceed $1$, and so $1\geq f_0\geq1-p_n$. Here, $f_0$ is chosen to maximize the objective, i.e., make $(1-p_n)f_0^{-1}\left|\bra{\phi_{0T}(0,0,0)}\hat{a}\ket{\phi_{0T}(0,0,0)}\right|^2$ as large as possible. Since $f_0$ is a quite complicated function of $n$ and the populations, we do not show its algebraic form here. Note that we make analogous omissions for $f_1$, $f_2$, etc., later. The ORT measure for states in this phase is:
\begin{equation}\label{eq:ORTMeasureNTriplet}
\begin{split}
\mathcal{N}(\hat{\rho}_{4F})&\lesssim3p_{n+3}+2p_{n+2}+p_{n+1}+n\\
&-(1-p_n)f_0^{-1}\left|\bra{\phi_{0T}(0,0,0)}\hat{a}\ket{\phi_{0T}(0,0,0)}\right|^2.
\end{split}
\end{equation}

A boundary on this phase is determined by $f_0=1-p_n$ such that $\ket{\phi_{0T}(\theta_j,\gamma_j,\zeta_j)}\rightarrow\ket{\phi_{Q}(\theta_j,\gamma_j,\zeta_j)}$ as $1-p_n\rightarrow f_0$. Interestingly, the condition $f_0\geq1-p_n$ is not sufficient to determine the entire boundary of this phase, because there are points $(p_{n+3},p_{n+2},p_{n+1})$ satisfying $f_0\geq1-p_n$ for which we determined an even better decomposition exists, hence the $(n+2,n+1)$-pair phase (see lower right panel of Fig. \ref{fig:sixPhases}). Ultimately, the $n$-triplet phase consists of the points satisfying $f_0\geq1-p_n$ that are not part of the $(n+2,n+1)$-pair phase.

Lastly, it is worth noting that the $p_{n+3}=0$ face of the pyramidal phase space of $\hat{\rho}_{4F}$ corresponds to the rank-$3$ case with populated photon numbers $n+2$, $n+1$, and $n$; the $n$-triplet phase is a smooth extension of the upper-pair phase on this rank-$3$ face.

\textit{$(n+1)$-triplet}--- The $(n+1)$-triplet phase is shown in the upper-right panel of Fig. \ref{fig:sixPhases}. Our ansatz for the optimal decomposition consists of one Fock state, $\ket{n+1}$, as well as states of the form:
\begin{small}
\begin{equation}\label{eq:phiOptimalnPlus1Triplet}
\begin{split}
&\ket{\phi_{1T}(\theta_j,\gamma_j,\zeta_j)}=\sqrt{1-f_1}e^{i\gamma_j}\ket{n+1}\\
&+\sqrt{f_1}\Bigg(\sqrt{\frac{p_{n+3}}{1-p_{n+1}}}\ket{n+3}+\sqrt{\frac{p_{n+2}}{1-p_{n+1}}}e^{i\theta_j}\ket{n+2}\\
&+\sqrt{\frac{p_{n}}{1-p_{n+1}}}e^{i\zeta_j}\ket{n}\Bigg).
\end{split}
\end{equation}
\end{small}
Such states contribute a total probability $(1-p_{n+1})f_1^{-1}$, which cannot exceed $1$, and so $1\geq f_1\geq1-p_{n+1}$. $f_1$ maximizes the objective, by making $(1-p_{n+1})f_1^{-1}\left|\bra{\phi_{1T}(0,0,0)}\hat{a}\ket{\phi_{1T}(0,0,0)}\right|^2$ as large as possible. The ORT measure for states in this phase is:
\begin{equation}\label{eq:ORTMeasureNPlus1Triplet}
\begin{split}
\mathcal{N}(\hat{\rho}_{4F})&\lesssim3p_{n+3}+2p_{n+2}+p_{n+1}+n\\
&-(1-p_{n+1})f_1^{-1}\left|\bra{\phi_{1T}(0,0,0)}\hat{a}\ket{\phi_{1T}(0,0,0)}\right|^2.
\end{split}
\end{equation}

A boundary on this phase is determined by $f_1=1-p_{n+1}$, where $\ket{\phi_{1T}(\theta_j,\gamma_j,\zeta_j)}\rightarrow\ket{\phi_{Q}(\theta_j,\gamma_j,\zeta_j)}$ as $1-p_{n+1}\rightarrow f_1$. As far as we determined from our numerical investigations, the $(n+1)$-triplet decomposition is optimal for all points $(p_{n+3},p_{n+2},p_{n+1})$ in $1\geq f_1\geq1-p_{n+1}$.

The $(n+1)$-triplet phase is a smooth extension of the upper-pair phase on the rank-$3$ face described by $p_{n+3}+p_{n+2}+p_{n+1}=1$.

\textit{$(n+2)$-triplet}--- The $(n+2)$-triplet phase is shown in the lower-left panel of Fig. \ref{fig:sixPhases}. The optimal decomposition consists of one Fock state, $\ket{n+2}$, as well as states of the form:
\begin{small}
\begin{equation}\label{eq:phiOptimalnPlus2Triplet}
\begin{split}
&\ket{\phi_{2T}(\theta_j,\gamma_j,\zeta_j)}=\sqrt{1-f_2}e^{i\theta_j}\ket{n+2}\\
&+\sqrt{f_2}\Bigg(\sqrt{\frac{p_{n+3}}{1-p_{n+2}}}\ket{n+3}+\sqrt{\frac{p_{n+1}}{1-p_{n+2}}}e^{i\gamma_j}\ket{n+1}\\
&+\sqrt{\frac{p_{n}}{1-p_{n+2}}}e^{i\zeta_j}\ket{n}\Bigg).
\end{split}
\end{equation}
\end{small}
Such states contribute a total probability $(1-p_{n+2})f_2^{-1}$, which cannot exceed $1$, and so $1\geq f_2\geq1-p_{n+2}$. $f_2$ maximizes the objective, by making $(1-p_{n+2})f_2^{-1}\left|\bra{\phi_{2T}(0,0,0)}\hat{a}\ket{\phi_{2T}(0,0,0)}\right|^2$ as large as possible. The ORT measure for states in this phase is:
\begin{equation}\label{eq:ORTMeasureNPlus2Triplet}
\begin{split}
\mathcal{N}(\hat{\rho}_{4F})&\lesssim3p_{n+3}+2p_{n+2}+p_{n+1}+n\\
&-(1-p_{n+2})f_2^{-1}\left|\bra{\phi_{2T}(0,0,0)}\hat{a}\ket{\phi_{2T}(0,0,0)}\right|^2.
\end{split}
\end{equation}

A boundary on this phase is determined by $f_2=1-p_{n+2}$, where $\ket{\phi_{2T}(\theta_j,\gamma_j,\zeta_j)}\rightarrow\ket{\phi_{Q}(\theta_j,\gamma_j,\zeta_j)}$ as $1-p_{n+2}\rightarrow f_2$. As far as we determined from our numerical investigations, the $(n+2)$-triplet decomposition is optimal for all points $(p_{n+3},p_{n+2},p_{n+1})$ in $1\geq f_2\geq1-p_{n+2}$.

The $(n+2)$-triplet phase is a smooth extension of the lower-pair phase on the rank-$3$ face described by $p_{n+3}=0$.

\textit{$(n+3)$-triplet}---The $(n+3)$-triplet phase is shown in the lower-middle panel of Fig. \ref{fig:sixPhases}. Our ansatz for the optimal decomposition consists of one Fock state, $\ket{n+3}$, as well as states of the form:
\begin{small}
\begin{equation}\label{eq:phiOptimalnPlus3Triplet}
\begin{split}
&\ket{\phi_{3T}(\theta_j,\gamma_j,\zeta_j)}=\sqrt{1-f_3}\ket{n+3}\\
&+\sqrt{f_3}\Bigg(\sqrt{\frac{p_{n+2}}{1-p_{n+3}}}e^{i\theta_j}\ket{n+2}+\sqrt{\frac{p_{n+1}}{1-p_{n+3}}}e^{i\gamma_j}\ket{n+1}\\
&+\sqrt{\frac{p_{n}}{1-p_{n+3}}}e^{i\zeta_j}\ket{n}\Bigg).
\end{split}
\end{equation}
\end{small}
Such states contribute a total probability $(1-p_{n+3})f_3^{-1}$, which cannot exceed $1$, and so $1\geq f_3\geq1-p_{n+3}$. Here, $f_3$ is chosen to maximize the objective, i.e., make $(1-p_{n+3})f_3^{-1}\left|\bra{\phi_{3T}(0,0,0)}\hat{a}\ket{\phi_{3T}(0,0,0)}\right|^2$ as large as possible. The ORT measure for states in this phase is:
\begin{equation}\label{eq:ORTMeasureNPlus3Triplet}
\begin{split}
\mathcal{N}(\hat{\rho}_{4F})&\lesssim3p_{n+3}+2p_{n+2}+p_{n+1}+n\\
&-(1-p_{n+3})f_3^{-1}\left|\bra{\phi_{3T}(0,0,0)}\hat{a}\ket{\phi_{3T}(0,0,0)}\right|^2.
\end{split}
\end{equation}

A boundary on this phase is determined by $f_3=1-p_{n+3}$ such that $\ket{\phi_{3T}(\theta_j,\gamma_j,\zeta_j)}\rightarrow\ket{\phi_{Q}(\theta_j,\gamma_j,\zeta_j)}$ as $1-p_{n+3}\rightarrow f_3$. The condition $f_3\geq1-p_{n+3}$, however, is not sufficient to determine the entire boundary of this phase, because there are points $(p_{n+3},p_{n+2},p_{n+1})$ satisfying $f_3\geq1-p_{n+3}$ for which {we determined an even better decomposition exists (such points belong to the $(n+2,n+1)$-pair phase). In fact, there are points satisfying both $f_3\geq1-p_{n+3}$ \textit{and} $f_0\geq1-p_n$, all of which belong to the $(n+2,n+1)$-pair phase.

The $p_{n+3}+p_{n+2}+p_{n+1}=1$ face of the pyramidal phase space of $\hat{\rho}_{4F}$ corresponds to the rank-$3$ case with populated photon numbers $n+3$, $n+2$, and $n+1$; the $(n+3)$-triplet phase is a smooth extension of the lower-pair phase on this rank-$3$ face.

\subsubsection{Pair Phase}
The $(n+2,n+1)$-pair phase is shown in the lower right panel of Fig. \ref{fig:sixPhases}. This phase is in surface contact with the $n$-triplet and $(n+3)$-triplet phases. Our ansatz for the optimal decomposition breaks $\hat{\rho}_{4F}$ into two parts: one part has support only for $\ket{n+3}$ and $\ket{n}$ (any decomposition is optimal for this part), while the other part is decomposed into states of the form:
\begin{small}
\begin{equation}\label{eq:phiOptimalnPlus2nPlus1Pair}
\begin{split}
&\ket{\phi_P(\theta_j,\gamma_j,\zeta_j)}=\sqrt{1-f}\left(\sqrt{g}\ket{n+3}+\sqrt{1-g}e^{i\zeta_j}\ket{n}\right)+\\
&\sqrt{f}\left(\sqrt{\frac{p_{n+2}}{p_{n+2}+p_{n+1}}}e^{i\theta_j}\ket{n+2}+\sqrt{\frac{p_{n+1}}{p_{n+2}+p_{n+1}}}e^{i\gamma_j}\ket{n+1}\right),
\end{split}
\end{equation}
\end{small}
where:
\begin{equation}\label{eq:nPlus2nPlus1PairOptimalG}
g=\frac{(3+n)p_{n+2}}{(1+n)p_{n+1}+(3+n)p_{n+2}}
\end{equation}
and $f=f(n,p_{n+2},p_{n+1})$ maximizes $(p_{n+2}+p_{n+1})f^{-1}|\bra{\phi_P(0,0,0)}\hat{a}\ket{\phi_P(0,0,0)}|^2$. These states contribute a total probability $(p_{n+2}+p_{n+1})f^{-1}$. Interestingly, $f$ and $g$ depend only on $n$, $p_{n+2}$, and $p_{n+1}$ (not on $p_{n+3}$ or $p_n$). Consequently, $(p_{n+2}+p_{n+1})f^{-1}|\bra{\phi_P(0,0,0)}\hat{a}\ket{\phi_P(0,0,0)}|^2$ depends only on $n$, $p_{n+2}$, and $p_{n+1}$. The ORT measure in this phase is:
\begin{equation}\label{eq:ORTMeasurenPlus2nPlus1Pair}
\begin{split}
\mathcal{N}(\hat{\rho}_{4F})&\lesssim3p_{n+3}+2p_{n+2}+p_{n+1}+n\\
&-(p_{n+2}+p_{n+1})f^{-1}\left|\bra{\phi_{P}(0,0,0)}\hat{a}\ket{\phi_{P}(0,0,0)}\right|^2.
\end{split}
\end{equation}
This regime is bounded by the inequalities:
\begin{subequations}\label{eq:nPlus2nPlus1PairBounds}
\begin{align}
p_{n+2} + p_{n+1} &\leq f\\
(1 - f)f^{-1}g(p_{n+2} + p_{n+1})&\leq p_{n+3}\\
(1 - f)f^{-1}(1-g)(p_{n+2} + p_{n+1})&\leq p_n.
\end{align}
\end{subequations}
Respectively, these inequalities ensure the $\ket{\phi_P(\theta_j,\gamma_j,\zeta_j)}$ states do not over-fill the total population, $\ket{n+3}$ population, and $\ket{n}$ population.

Lastly, we note that the $(n+2,n+1)$-pair phase may be regarded as a further refinement of the $n$-triplet and $(n+3)$-triplet phases. It lies within the union of $f_0\geq1-p_{n}$ and $f_3\geq1-p_{n+3}$ and covers the intersection of $f_0\geq1-p_{n}$ and $f_3\geq1-p_{n+3}$. Within its domain of validity, the $(n+2,n+1)$-pair decomposition is superior to that of the corresponding triplet phases.

\subsubsection{Exceptions}\label{subsec:exceptions}
As previously stated, the reader should regard the ansatzes for the ORT measure here as close and useful upper bounds, and these phases as a useful guide, akin to a cartographer's initial map. We can only check so many points within the full pyramidal phase space, so it is challenging to arrive at a complete and exact picture. In our checks of these ansatzes, we \textit{did} happen to discover examples wherein the nonclassicality surpasses the bounds presented here, albeit by small margins (producing no distinguishable effect on Fig. \ref{fig:sixPhases}). We provide examples here for the sake of transparency.

One state that surpasses the aforementioned bounds is $0.01\ket{3}\bra{3}+0.01\ket{2}\bra{2}+0.06\ket{1}\bra{1}+0.92\ket{0}\bra{0}$. Our ansatz placed this point in the $n$-triplet phase ($n=0$ here), and Eq. ~\eqref{eq:ORTMeasureNTriplet} bounds its nonclassicality as $\mathcal{N}\leq0.01625$. Linear optimization with a binning of $\Delta=.0099$\footnote{With this choice of $\Delta$, the optimized vector has dimension 537052.} bounds the nonclassicality as $\mathcal{N}\lesssim0.0153096$, about a $6\%$ difference. This case was chosen for its relatively high disparity, so that the $n$-triplet ansatz is still within about $6\%$ may be interpreted as a point in its favor. It is worth comparing to the lower bound given by the metrological power $\mathcal{W}$. Interestingly, for this state, $\mathcal{W}=0$ --- it has zero metrological advantage over coherent states for quadrature sensing. This also shows that the $n$-triplet ansatz is a more useful bound than $\mathcal{W}$.

Another counterexample to our ansatzes is $0.01\ket{3}\bra{3}+0.01\ket{2}\bra{2}+0.15\ket{1}\bra{1}+0.83\ket{0}\bra{0}$. Our ansatz placed this point in the quartet phase and Eq. ~\eqref{eq:ORTMeasureQuartet} bounds its nonclassicality at $\mathcal{N}\leq0.0194274$, whereas linear programming with a $\Delta=.0099$ yielded a value of $\mathcal{N}\lesssim0.0190649$, about a $2\%$ difference. Based on this result, we conclude that this state is not actually a part of the quartet phase (i.e., we were slightly incorrect about the quartet phase's boundary). We remain confident that the quartet phase exists, (i.e., that there are rank-4 Fock-diagonal states that saturate Eq. ~\eqref{eq:simpleBound}), although we were incorrect about the precise boundary. Metrological power $\mathcal{W}=0$ for this state as well.

It is worth noting these two states, in addition to having zero metrological power, are close to the surface border of the $n$-triplet and quartet phases (see Fig. \ref{fig:sixPhases}, noting that the quartet phase has a lower surface that does not show from the viewing angle); they are not isolated exceptions, but rather part of a continuous section (or sections), the boundaries of which we do not know. While one may hope for a more complete and exact picture of the true optimal phases in the rank-4 case, we believe that any correct ansatz will not only be complicated but also impossible to verify with the methods here. Thus, we suggest linear programming for evaluating nonclassicality in specific cases, while taking the phases presented here as a useful analytical guide.

\subsection{Higher-rank cases}
To demonstrate the scope of our linear programming numerical method, we apply it to higher-rank mixed Fock states of the form given by Eq.~\eqref{eq:MFock}. As an example, we consider truncated thermal states, where the populations of $\ket{n}$ $(n\ge M)$ are reduced to zero:
\begin{equation}\label{eq:thermalRho}
\hat{\rho}_{MF}=\sum_{k=0}^{M-1}p_{k}\ket{k}\bra{k},
\end{equation}
where the populations are:
\begin{equation}\label{eq:thermalRhoPopulations}
p_k=\mathcal{A}\frac{n_{\text{th}}^k}{(1+n_{\text{th}})^{k+1}}
\end{equation}
with $n_{\text{th}}$ the mean photon number of the field before the truncation and:
\begin{equation}\label{eq:thermalRhoNormalization}
    \mathcal{A}=\frac{1}{1-\left(\frac{n_{\text{th}}}{1+n_{\text{th}}}\right)^M}
\end{equation}
is the normalization factor after truncating the thermal states. Even though true (i.e., untruncated) thermal states are diagonal in the Fock basis, they are classical, meaning $\mathcal{N}=0$ and their $P$-function is positive. This classicality is only possible because their expansion in terms of Fock states has infinitely many terms, which can be equivalently represented by a mixture of coherent states~\cite{SZ}.
\begin{figure}
    \centering
    \includegraphics[width=0.45\textwidth]{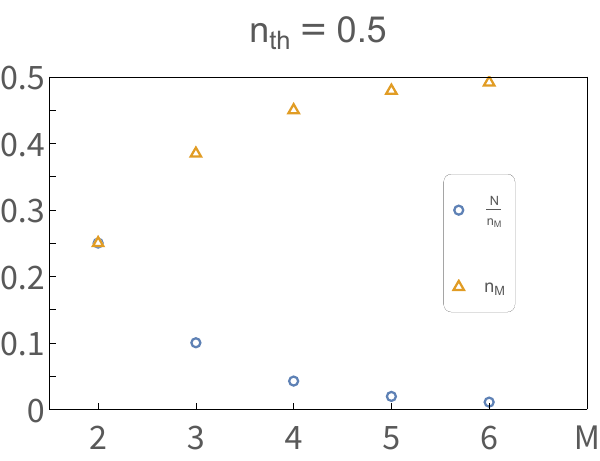}
    \caption{Nonclassicality per unit average energy, $\mathcal{N}/n_M$, along with average energy $n_M$, of truncated thermal states. The original thermal state has average energy $n_{th}=0.5$ and the truncated thermal states are obtained by projecting out all photon numbers greater than or equal to $M$. The truncated thermal states should resemble the original thermal state (which has nonclassicality $0$) more as $M$ increases, and indeed we see $\mathcal{N}/n_M\rightarrow0$ as $M$ increases.}
    \label{fig:truncatedThermal}
\end{figure}

 Truncated thermal states should have nonzero nonclassicality. However, increasing the number of terms in a truncated thermal state should cause the nonclassicality to approach zero (i.e., recover the thermal state's classicality). We demonstrate this in Fig. \ref{fig:truncatedThermal}, where we evaluate the ORT measure, per unit average energy, using our numerical method in the rank-5 and rank-6 cases. We choose a low $n_{\text{th}}=0.5$ such that the expected photon number after truncation at $M=6$ is close to $n_{\text{th}}$: $n_6=0.491758$. Nonclassicality per unit energy $\mathcal{N}/n_M$ indeed decreases from its $M=2$ value of $0.25$ to a $M=6$ value of $\lesssim .011$.

\section{Conclusion}
In this work, we introduced a method, based on linear programming, for evaluating the ORT measure of Fock-diagonal mixed states. In doing so, we have further advanced the ORT measure as a useful tool for characterizing bosonic nonclassicality. Our work also suggests that linear programming may be useful for solving convex roof optimization problems in other contexts, such as entanglement quantification.

We have provided in-depth results for the case where three or four neighboring Fock states are populated. Analytical results were obtained by carefully analyzing the optimal histograms given by the linear programming method. We found that, depending on the populations, Fock-diagonal states can be in various different phases, wherein different types of decompositions are optimal (for ORT measure purposes). We also noted that optimal decompositions fall into two categories: ``simple" and ``composite," a distinction which may be useful for solving other convex roof optimization problems.

We gave some numerical results for rank-$5$ and rank-$6$ states, showing that the method remains applicable. As the rank increases, the computational complexity does increase, as more bins are needed for the optimal histogram to be accurate. However, we suspect that such a challenge can be met by employing an iterative, adaptive approach.

Since the ORT measure is resource-theoretic, and related to metrological power, our results help facilitate a deeper understanding of the resources behind quantum-enhanced sensing. Moving forward, we would like to calculate the nonclassicality of mixed states that have coherences in the Fock basis. Furthermore, we would like to apply our results to study the tradeoff between entanglement generation and individual mode nonclassicality in a beam-splitter~\cite{vogel2014unified, ge2015conservation,Arkhipov:2016aa, liu2023classical}. 

\section*{Acknowledgements}
We thank Luke Ellert-Beck and Garrett Jepson for helpful discussions.

This research is supported by NSF Award 2243591.

\appendix
\section{Optimizing pair phase coefficients}\label{appendixA}
To see that the value of $f$ in Equation \eqref{eq:optimalF} is optimal, note that the goal is to maximize $(p_{n+2}+p_{n+1})f^{-1}\left|\bra{\phi_U(0,0}\hat{a}\ket{\phi_U(0,0}\right|^2$, where $(p_{n+2}+p_{n+1})f^{-1}$ is the total probability invested in states of the form $\ket{\phi_U(\theta_j,\gamma_j}$ (see Eq. \eqref{eq:ORTFockDiagonalRewrite}). The upper-pair decomposition also includes the lowest populated Fock state $\ket{n}$, but $\bra{n}\hat{a}\ket{n}=0$ and hence this state does not contribute to the summation in Eq. \eqref{eq:ORTFockDiagonalRewrite}.

Since:
\begin{equation}
\begin{split}
&\left|\bra{\phi_U(0,0}\hat{a}\ket{\phi_U(0,0}\right|^2=\\
&f\frac{p_{n+1}}{p_{n+2}+p_{n+1}}\left(\sqrt{\frac{(n+2)p_{n+2}}{p_{n+2}+p_{n+1}}}\sqrt{f}+\sqrt{n+1}\sqrt{1-f}\right)^2,
\end{split}
\end{equation}
we have:
\begin{equation}
\begin{split}
&(p_{n+2}+p_{n+1})f^{-1}\left|\bra{\phi_U(0,0}\hat{a}\ket{\phi_U(0,0}\right|^2=\\
&p_{n+1}\left(\sqrt{\frac{(n+2)p_{n+2}}{p_{n+2}+p_{n+1}}}\sqrt{f}+\sqrt{n+1}\sqrt{1-f}\right)^2.
\end{split}
\end{equation}
We maximize this with respect to $f$, subject to probability constraint $p_{n+2}+p_{n+1}\leq f\leq1$. Solving:
\begin{equation}
\frac{\partial}{\partial f}\left(\sqrt{\frac{(n+2)p_{n+2}}{p_{n+2}+p_{n+1}}}\sqrt{f}+\sqrt{n+1}\sqrt{1-f}\right)^2=0,
\end{equation}
leads to the equation:
\begin{equation}
\frac{(n+2)p_{n+2}}{p_{n+2}+p_{n+1}}f^{-1}=(n+1)(1-f)^{-1},
\end{equation}
the solution of which is given in Eq. \eqref{eq:optimalF}. One may verify that this value of $f$ yields the maximum over the space $p_{n+2}+p_{n+1}\leq f\leq 1$.

\section{Optimizing phase-dependence}\label{appendixB}
In subsection \ref{subsec:higherRankFock},we claimed that, given any decomposition $\{q(\vec{x},\vec{\theta}),\ket{\phi\left(\vec{x},\vec{\theta}\right)}=\sum_{k=0}^{M-1}x_ke^{i\theta_k}\ket{n+k}\}$ of $\hat{\rho}_{MF}=\sum_{k=0}^{M-1}p_{n+k}\ket{n+k}\bra{n+k}$, one may always construct an alternative decomposition $\tilde{q}(\vec{x},\vec{\theta})$ (with modified $\vec{\theta}$-dependence) that is at least as optimal as the originating $q(\vec{x},\vec{\theta})$, thus reducing the ORT measure calculation to an optimization over $\mathcal{Q}(\vec{x})\equiv\sum_{\vec{\theta}}\tilde{q}(\vec{x},\vec{\theta})$ . We justify that claim here.

It is sufficient to demonstrate that we can construct a $\tilde{q}(\vec{x},\vec{\theta})$ that: (a) satisfies $\sum_{\vec{\theta}}\tilde{q}(\vec{x},\vec{\theta})|\bar{\alpha}_{\vec{x},\vec{\theta}}|^2\geq\sum_{\vec{\theta}}q(\vec{x},\vec{\theta})|\bar{\alpha}_{\vec{x},\vec{\theta}}|^2$ for all $\vec{x}$, (b) satisfies $\sum_{\vec{\theta}}\tilde{q}(\vec{x},\vec{\theta})\bar{\alpha}_{\vec{x},\vec{\theta}}^2=0$ for all $\vec{x}$, and (c) is also a valid decomposition. $\tilde{q}(\vec{x},\vec{\theta})$'s superiority to $q(\vec{x},\vec{\theta})$ is then clear by reference to Eq. \eqref{eq:ORTFockDiagonal}.

Our proof is by example. To avoid complicated $\delta$-function expressions, we write $\tilde{q}$ as $\tilde{q}_j(\vec{x})$, where $j=0, 1, 2, ..., P-1$ is an integer playing the role of $\delta$-functions in $\vec{\theta}$. We decompose $\hat{\rho}_{MF}$ as $\hat{\rho}_{MF}=\sum_{\vec{x}}\sum_j\tilde{q}_j(\vec{x})\ket{\phi_j(\vec{x})}\bra{\phi_j(\vec{x})}$
where:
\begin{subequations}
\begin{equation}
\tilde{q}_j(\vec{x}) = \frac{1}{P}\left(\sum_{\vec{\theta}'}q(\vec{x},\vec{\theta}')\right)\\
\end{equation}
\begin{equation}
\ket{\phi_j(\vec{x})}=\sum_{k=0}^{M-1} x_k e^{i\frac{2\pi j}{P}(M-1-k)}\ket{n+k}.
\end{equation}
\end{subequations}
Notably,
\begin{equation}
\begin{split}
\bar{\alpha}_{j,\vec{x}}&=\bra{\phi_j(\vec{x})}\hat{a}\ket{\phi_j(\vec{x})}\\
&=\left(\sum_{k=0}^{M-2} x_{k+1}x_k\sqrt{n+k+1}\right)e^{-i\frac{2\pi j}{P}}.
\end{split}
\end{equation}
The size of $\bar{\alpha}_{j,\vec{x}}$ is maximal (for a given $\vec{x}$), since the terms in the summation constructively interfere. Thus property (a), $\sum_{j}\tilde{q}_j(\vec{x})|\bar{\alpha}_{j,\vec{x}}|^2\geq\sum_{\vec{\theta}}q(\vec{x},\vec{\theta})|\bar{\alpha}_{\vec{x},\vec{\theta}}|^2$, holds as desired.

Since $|\bar{\alpha}_{j,\vec{x}}|$ is independent of $j$:
\begin{equation}
\sum_{j=0}^{P-1}\tilde{q}_j(\vec{x})\bar{\alpha}_{j,\vec{x}}^2\propto\sum_{j=0}^{P-1} \left(e^{-i\frac{2\pi j}{P}}\right)^2=0.
\end{equation}
Thus, property (b) holds as desired. Here, we are assuming that $P\geq3$. This result relies on the fact that the sum of the squares of the $P$-th roots of unity is zero.

To demonstrate property (c), we use the more general result that if $\epsilon=e^{i\frac{2\pi}{P}}$, and $l<P$ is a positive integer, then $\sum_{j=0}^{P-1}(\epsilon^{j})^l=\frac{1-\epsilon^{Pl}}{1-\epsilon^l}=0$. $\hat{\rho}_{MF}$ has zero coherences in the Fock basis, and indeed for $k\neq k'$ (between $0$ and $M-1$ inclusive):
\begin{equation}
\begin{split}
&\bra{n+k}\left(\sum_{j=0}^{P-1}\tilde{q}_j(\vec{x})\ket{\phi_j(\vec{x})}\bra{\phi_j(\vec{x})}\right)\ket{n+k'}\\
&\propto\sum_{j=0}^{P-1}\left(e^{i\frac{2\pi j}{P}}\right)^{k'-k}=0,
\end{split}
\end{equation}
so long as $P\geq M$. Thus, in all, we should have the integer $P\geq\max(3,M)$.

That $\tilde{q}_j(\vec{x})$ also satisfies the population constraints for $\hat{\rho}_{MF}$ is a simple consequence of the fact that the photon number populations depend only on the ``marginal'' $\vec{x}$ distributions, and $\sum_j\tilde{q}_j(\vec{x})=\sum_{\vec{\theta}}q(\vec{x},\vec{\theta})$. $q(\vec{x},\vec{\theta})$ satisfies the population constraints, so $\tilde{q}$ must as well. This concludes our proof.

The last thing we would like to note here is that 
\begin{equation}
\frac{1}{\sum_{\vec{\theta}'}q(\vec{x},\vec{\theta}')}\sum_{j=0}^{P-1}\tilde{q}_j(\vec{x})\ket{\phi_j(\vec{x})}\bra{\phi_j(\vec{x})}=\hat{\rho}_{\vec{x}}
\end{equation}
where $\hat{\rho}_{\vec{x}}=\sum_{k=0}^{M-1}x_k^2\ket{n+k}\bra{n+k}$ is a normalized Fock-diagonal state. This is Equation ~\eqref{eq:tildeQHigherRank}. The purpose of the linear program, Eq. ~\eqref{eq:ORTDiscreteMF}, is to optimize the marginal $\vec{x}$ distribution $\mathcal{Q}(\vec{x})$ subject to the constraint $\hat{\rho}_{MF}=\sum_{\vec{x}}\mathcal{Q}(\vec{x})\hat{\rho}_{\vec{x}}$.

\bibliography{ref.bib}

\end{document}